\documentclass[prb,twocolumn,showpacs,amsmath,amssymb]{revtex4-1}
\usepackage{graphicx}
\usepackage{dcolumn}
\usepackage{bm}
\usepackage{amsmath}
\DeclareGraphicsExtensions{.pdf,.eps,.png,.jpg,.mps}

\usepackage{graphicx}
\usepackage{dcolumn}
\usepackage{bm}

\begin{document}

\preprint{APS/123-QED}

\title{Fano-Andreev effect in T-shape double-quantum-dot in the Kondo regime}

\author{A. M. Calle}
\affiliation{%
 Departamento de F\'isica, Universidad T\'ecnica Federico Santa Mar\'ia, Avenida Espa\~{n}a 1680, Casilla 110V, Valpara\'iso, Chile \\
}
\author{M. Pacheco}
\affiliation{%
 Departamento de F\'isica, Universidad T\'ecnica Federico Santa Mar\'ia, Avenida Espa\~{n}a 1680, Casilla 110V, Valpara\'iso, Chile \\
}

\author{G. B. Martins}
\affiliation{
 Department of Physics, Oakland University, Rochester, MI 48302, USA
}
\affiliation{Instituto de F\'{\i}sica, Universidade Federal Fluminense, 24210-346 Niter\'oi, RJ, Brazil\\}
\author{V. M. Apel}
\affiliation{
 Departamento de F\'isica, Universidad Cat\'olica del Norte, Angamos 0610, Casilla 1280, Antofagasta, Chile\\
}

\author{G. A. Lara}
\affiliation{
 Departamento de F\'isica, Universidad de Antofagasta, Casilla 170, Antofagasta, Chile\\
}
\author{P. A. Orellana}%
 \email{pedro.orellana@usm.cl}
\affiliation{%
 Departamento de F\'isica, Universidad T\'ecnica Federico Santa Mar\'ia, Avenida Espa\~{n}a 1680, Casilla 110V, Valpara\'iso, Chile \\
}
\date{\today}

\begin{abstract}
	In the present work, we investigate the electronic transport through a T-shape double quantum dot system coupled to two normal leads and to one superconducting lead. We explore the interplay between Kondo and Andreev states due to proximity effects. We find that Kondo resonance is modified by the Andreev bound states, which manifest through Fano antiresonances in the  local density of states of the embedded quantum dot and normal transmission. This means that there is a correlation between Andreev bound states and Fano resonances that is robust under the influence of high electronic correlation. We have also found that the dominant couplings at the quantum dots are characterized by a crossover region that defines the range where the Fano-Kondo and the Andreev-Kondo effect prevail in each quantum dot. Likewise, we find that the interaction between Kondo and Andreev bound states has a notable influence on the Andreev transport.
\end{abstract}

\maketitle

\section{\label{sec:level1} Introduction}

Superconductivity is a macroscopic quantum phenomenon involving large number of electrons \cite{Tinkham}. As described by Bardeen, Cooper, and Schrieffer (BCS) \cite{BCS} electrons in condensed matter with an attractive interaction condense into a superconducting state below a critical temperature, referred to as the BCS state. In this state, electrons with antiparallel spins form singlet bound states (S = 0) known as Cooper pairs. This pair formation is a fermionic many-body phenomenon as it relies on the existence of a Fermi surface. In contrast, electrons in the normal (N) phase of metals behave very differently to those in the superconducting phase. For example, it is possible to trap small numbers of electrons in sub-micrometer-sized boxes known as quantum dots, which are systems in which electrons are confined in all space dimensions and as a consequence of this confinement energy and charge are quantized. \cite{Kouwenhoven, Jacak} This confinement can, under appropriate conditions, generate highly correlated ground states, like the Kondo state.

Fascinating physical properties and interesting device applications arise when the ability to control single electrons in quantum dots is linked with superconductivity. 
One of these properties is the so-called proximity effect, whose most important characteristic is the Andreev reflection, which takes place in a normal-metal/superconductor interface (NS). In Andreev reflections \cite{Andreev}, an electron, in the normal metal side, with an energy in the superconducting gap is reflected at the interface as a hole. The corresponding charge $2e$ is transferred to the Cooper pair which appears on the superconducting side of the interface \cite{Review_Buzdin}. Hence, the single electron states of the normal metal are converted into Cooper pairs in the superconductor [Fig.~\ref{system}(b)]. Andreev reflection plays an important role for the understanding of quantum transport properties of NS systems. Moreover, in zero dimensional structures, as quantum dots, this process can give rise to discrete entangled electron-hole states confined to the quantum dot, called Andreev bound states (ABS). \cite{Pillet} Since it was realized that Majorana fermions may be used for fault-tolerant quantum computation, there has been renewed interest in Andreev bound states, as Majorana fermions are zero-energy Andreev bound states that exist at the surface (or in a vortex core) of a topological superconductor. \cite{Hasan, Beenakker2, Sato}

A singlet ground state due to many-body effects also occurs in a quite different situation, when a magnetic impurity is embedded in a metallic host. \cite{Kondo, Hewson} This state, known as a Kondo singlet, occurs because the electrons in the metal at low temperature experience a large effective coupling to the localized impurity spin. As a consequence, it is energetically favorable to screen the local moment, resulting in a (Kondo) singlet state (S = 0) \cite{Bauer} universally characterized by the Kondo temperature $T_{K}$. The Kondo effect produces a signature in the electronic spectral density in the form of a resonance peak at the Fermi energy. \cite{Revival_Kondo}
The competition between Kondo and Fano effect \cite{Fano} has also been investigated. \cite{Chung, Tanaka2} In this case the Kondo resonance line shape is appreciably modified by the Fano effect. In particular, side coupled double quantum dots have been widely reviewed \cite{Zitko, Sasaki, Kormanyos} since it allows the study of the interference between two transmission channels, a resonant and a non-resonant one. The study of this kind of systems reveals an interesting interplay between many-particle effects and quantum interference.

On the other hand, BCS superconductivity and the Kondo effect have been extensively studied. The interaction and competition of these two effects in hybrid superconductor nanostructures have attracted a lot of interest lately. The interplay between the Kondo effect and superconductivity in normal metal/quantum dot/superconductor (N-QD-S) devices, has been investigated recently because the subgap transport shows very rich features. \cite{Deacon, Deacon2, Eichler, Domanski, Li,Koerting, Yamada}
A number of works have addressed the problem of an Anderson impurity coupled to a single superconductor, either by numerical renormalization group (NRG) calculations \cite{Bauer} or auxiliary-boson methods \cite{Clerk} and have explored the intricate competition between Cooper-pairing and local correlations. \cite{Yeyati} 

In the present paper we study the electronic transport through a system of two quantum dots coupled in T-shape geometry to two metallic leads and one superconducting (SC) lead [Fig.~\ref{system}(a)]  
The study of this setup is encouraged by the assumption that the T-shape double quantum dot system will be a good system to study the interplay between the Kondo physics, Andreev bound states and Fano effect. In fact, we show that the Kondo resonance is modified by the Andreev bound states, which manifest through Fano antiresonances in the local density of states of the embedded quantum dot and normal transmission. Besides, we find that the system shows a crossover region, 
when coupling to the SC lead is varied, where the Fano-Kondo and the Andreev-Kondo effect prevail in each quantum dot. 
Previous work, related to ours, includes Ref.~\onlinecite{Baranski2011}, where the authors study a 
double dot in T-shape geometry sandwiched between normal and SC leads.

The paper is organized as follows. In Sec. II we describe the model to study the $(L, R)-QD2-QD1-S$ system. We also outline the mean field slave bosons approach, as well as the theoretical framework based on the non-equilibrium Green's function techniques. In Sec. III, we discuss the numerical results obtained and, finally, a brief summary is given.

\section{Description of the Model}

\begin{figure}[ht]
\centerline{\includegraphics[width=80mm,height=38mm,clip]{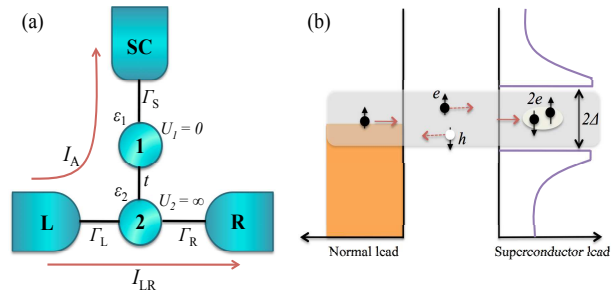}}
	\caption{a) Schematic view of T-shape DQD system coupled to left ($L$) and right ($R$) normal leads and an SC lead ($S$) with an interdot coupling denoted by $t$. b) Andreev reflection: the electron (e) is reflected as a hole (h) with the same momentum
and opposite velocity. The missing charge of $2e$ is absorbed as a Cooper pair by the superconductor.}
\label{system}
\end{figure}

In this paper we consider a system composed by a T-shape double quantum dot, with a  single-level in each quantum dot, which are coupled to two normal metallic leads and to an SC lead, $(L, R)-QD2-QD1-S$, as shown in Fig.~\ref{system}(a). We consider a strong intradot Coulomb interaction $U$ in the embedded quantum dot QD2. In our model the double quantum dot is modeled by a two impurity Anderson Hamiltonian and the Hamiltonian for the whole system can be written as:

\begin{equation}
\label{1}
H=H_{L}+H_{R}+H_{S}+H_{dot}+H_{T} \hspace{0.1cm}.
\end{equation}

\noindent where $H_{L(R)}$ is the Hamiltonian for the left (right) normal lead, which is given by
\begin{equation}
\label{2}
H_{L(R)}=\sum_{k_{L(R)}\sigma}\epsilon_{k_{L(R)}}C^{\dag}_{k_{L(R)}\sigma}C_{k_{L(R)}\sigma} \hspace{0.1cm},
\end{equation}
\noindent being $C^{\dag}_{k_{L(R)}\sigma}$ and $C_{k_{L(R)}\sigma}$ creation and annihilation operator for electrons with momentum $k_{L(R)}$ and spin $\sigma$ in the metallic lead $L(R)$, respectively.  The standard BCS Hamiltonian for the SC lead is
\begin{equation}
\label{3}
H_{S}=\sum_{k_{S}\sigma}\epsilon_{k_{S}}C^{\dag}_{k_{S}\sigma}C_{k_{S}\sigma}+\sum_{k_{S}}\Delta\left( C^{\dag}_{k_{S}\uparrow}C^{\dag}_{-k_{S}\downarrow}+h.c.\right)\hspace{0.1cm},
\end{equation}
\noindent where $C^{\dag}_{k_{S}\sigma}$ and $C_{k_{S}\sigma}$ are the creation and annihilation operators for electrons in the SC lead, while $\Delta$ is the superconducting gap function which is assumed to be s-wave, i.e., $k$-independent and real ($\Delta^{\dag}=\Delta$). The Hamiltonian for the double quantum dot is given by
\begin{eqnarray}
\label{4 }
H_{dot}&=&\sum_{i\sigma}\epsilon_{i}d^{\dag}_{i\sigma}d_{i\sigma}+U \hspace{0.05cm} n_{2\uparrow}n_{2\downarrow}\hspace{0.1cm},
\end{eqnarray}
\noindent $d^{\dag}_{i\sigma}$($d_{i\sigma}$) being the creation (annihilation) operator for electrons in the quantum dot level $\epsilon_{i}$ ($i=1,2$). $U_{2}\equiv U$ is the intradot Coulomb interaction in the QD2 and we set $U_{1}=0$.  Finally, the tunneling between the QD's and leads is described by 
\begin{eqnarray}
\label{5}
H_{T}&=&\sum_{\sigma}t\left(d^{\dag}_{1\sigma}d_{2\sigma}+d^{\dag}_{2\sigma}d_{1\sigma}\right)\nonumber\\&+&
\sum_{k_{L(R)}\sigma}\left(V_{k_{L(R)}}C^{\dag}_{k_{L(R)}\sigma}d_{2\sigma}+h.c.\right)\nonumber\\&+&
\sum_{k_{S}\sigma}\left(V_{k_{S}}C^{\dag}_{k_{S}\sigma}d_{1\sigma}+h.c.\right)\hspace{0.1cm}.
\end{eqnarray}

\noindent The embedded quantum dot (QD2) is coupled to the side-coupled quantum dot (QD1) via the interdot coupling $t$, which is taken as being a real parameter. $V_{k_{L(R)}}$ and $V_{k_{S}}$ are the coupling between left (right) normal lead and QD2 and between SC lead and QD1, respectively.
We consider the density of states describing the left (right) lead $\rho_{L(R)}$ as being constant and equal to $1/D$, where $D$ is the lead bandwidth. The coupling strength between QD2 and the leads is given by $\Gamma_{L(R)}=2\pi\rho_{L(R)}V^{2}_{k_{L(R)}}$.

In the following analysis, we consider $U$ sufficiently large ($U$ $\rightarrow \infty$) so that the double occupancy in QD2 is forbidden. Hence, we will employ a simple mean field approach known as slave-boson mean field approximation (SBMFA). This method was introduced by Coleman for the Anderson model \cite{Coleman} and is based on the introduction of auxiliary boson fields $b^{\dag}_{0}$, $b_{0}$, which act as projector onto the empty impurity state. In addition, $f^{\dag}_{2\sigma}$ ($f_{2\sigma}$) is the creation (annihilation) operator for pseudo-fermions introduced to describe singly occupied states in QD2. Therefore, the creation (annihilation) operator for electrons in QD2, $d^{\dag}_{2\sigma}$ ($d_{2\sigma}$) is then replaced by $f^{\dag}_{2\sigma}b_{2}$ ($b^{\dag}_{2}f_{2\sigma}$). 
In order to avoid double occupation, these operators should satisfy the completeness relation
$b^{\dag}_{2}b_{2}+\sum f^{\dag}_{2\sigma}f_{2\sigma}=1$.

The mean field approximation (MFA) is based on replacing the boson field $b_{2}$ and $b^{\dag}_{2}$ by their expectation values $\langle b^{\dag}_{2} \rangle= \langle b_{2}\rangle \equiv \widetilde{b}_{2}$ in the Hamiltonian. Hence, introducing the renormalized parameters $\widetilde{\epsilon}_{2}=\epsilon_{2}+\lambda $, $\widetilde{t}=\widetilde{b}_{2}\hspace{0.05cm}t$ and $\widetilde{V}_{k_{L(R)}}=\widetilde{b}_{2}V_{k_{L(R)}}$ the mean field Hamiltonian can be written as

\begin{eqnarray}
\label{Hmfa}
H_{MFA}&=&
\sum_{k_{L(R)}\sigma}\left( \epsilon_{k_{L(R)}\sigma}C^{\dag}_{k_{L(R)}\sigma}C_{k_{L(R)}\sigma} \right)\nonumber\\&+&
\sum_{k_{S}\sigma}\left( \epsilon_{k_{S}\sigma}C^{\dag}_{k_{S}\sigma}C_{k_{S}\sigma} \right)+
\sum_{k_{S}}\left(\Delta \hspace{0.05cm} C^{\dag}_{k_{S}\uparrow}C^{\dag}_{-k_{S}\downarrow}+h.c\right)\nonumber\\&+&
\sum_{\sigma}\left( \widetilde{\epsilon}_{2}f^{\dag}_{2\sigma}f_{2\sigma}+\epsilon_{1}d^{\dag}_{1\sigma}d_{1\sigma} \right)+
\sum_{\sigma}\left(\widetilde{t}  \hspace{0.05cm} d^{\dag}_{1\sigma}f_{2\sigma}+h.c\right)\nonumber\\&+&
\sum_{k_{L(R)}\sigma}\left( \widetilde{V}_{k_{L(R)}}C^{\dag}_{k_{L(R)}\sigma}f_{2\sigma}+h.c \right)\nonumber\\&+&
\sum_{k_{S}\sigma}\left(V_{k_{S}}C^{\dag}_{k_{S}\sigma}d_{1\sigma}+h.c \right)+
\lambda\left( b^{2}_{2}-1 \right) \hspace{0.1cm},
\end{eqnarray}

\noindent where $H_{MFA}$ is a single particle Hamiltonian which depends of the unknown parameters $\widetilde{b}_{2}$ and $\lambda$. These parameters can be determined by minimizing the ground state energy of the effective MFA Hamiltonian with respect to $\widetilde{b}_{2}$ and $\lambda$. Conditions for minimal energy together with the application of the Hellmann-Feynman theorem $\left(\frac{\partial}{\partial x}\langle H\rangle=\langle\frac{\partial}{\partial x}H\rangle\right)$ to the Hamiltonian, gives a set of self-consistent equations, which can be written in terms of the lesser Green's functions as follows \cite{Trocha-3QDS}

\begin{equation}
\label{selfc1}
-\sum_{\sigma}\dot{\imath}\int^{\infty}_{-\infty} \frac{d\omega}{2\pi } \langle\langle f_{2\sigma},f^{\dag}_{2\sigma}\rangle\rangle^{<}_{\omega}+\widetilde{b}^{2}_{2}=1 \hspace{0.1cm},
\end{equation}

\begin{equation}
\label{selfc2}
-\sum_{\sigma}\dot{\imath}\int^{\infty}_{-\infty} \frac{d\omega}{2\pi}\left( \omega-\widetilde{\epsilon}_{2}\right) \langle\langle f_{2\sigma},f^{\dag}_{2\sigma} \rangle\rangle^{<}_{\omega}+\lambda \hspace{0.05cm}  \widetilde{b}^{2}_{2}=0 \hspace{0.1cm}.
\end{equation}

When QD1 is decoupled from QD2, the Kondo temperature of QD2 at equilibrium is given by $T^{0}_{K}=De^{-\pi|\epsilon_{2}-E_{F}|/\Gamma}$, where $\Gamma=\Gamma_{L}+\Gamma_{R}$. The self-consistently determined parameters $\widetilde{\epsilon}_{2}$ and $\widetilde{\Gamma}_{L(R)}=\widetilde{b}^{2}_{2}\Gamma_{L(R)}$ give the position and the width of the Kondo peak in QD2. \cite{Wu} To solve (\ref{selfc1}) and (\ref{selfc2}) we still need to determine the lesser Green's function $ \langle\langle f_{2\sigma},f^{\dag}_{2\sigma}\rangle\rangle^{<}_{\omega}$. In order to obtain the Green's functions for the system, we use the Equation of Motion (EOM) method, in which the Green's functions can be written in a compact matrix form as the Dyson equation
$\boldsymbol{G}^{r}_{j,\sigma}=\boldsymbol{g}^{r}_{j,\sigma}+\boldsymbol{g}^{r}_{j,\sigma} \hspace{0.1cm} \boldsymbol{\Sigma}^{r}_{j} \hspace{0.1cm} \boldsymbol{G}^{r}_{j,\sigma}$,
where $\boldsymbol{g}^{r}_{j,\sigma}$ is the Green's functions for a non-interacting QD and $\boldsymbol{\Sigma}^{r}$ the retarded self-energy. At this point is useful introduce the Nambu spinor notation, in which the retarded and lesser Green's functions can be written as \cite{Nambu} 
\begin{equation}
\label{Nambur}
\boldsymbol{G}^{r}(t,t')=-\dot{\imath}\theta(t-t')\langle\boldsymbol{\Psi}(t),\boldsymbol{\Psi}^{\dag}(t')\rangle \hspace{0.1cm},
\end{equation}

\begin{equation}
\label{Nambul}
\boldsymbol{G}^{<}\left(t,t'\right)=\dot{\imath}\langle\boldsymbol{\Psi}^{\dag}\left(t'\right) \hspace{0.1cm} \boldsymbol{\Psi}\left(t\right)\rangle \hspace{0.1cm},
\end{equation}

\noindent where $\boldsymbol{\Psi}^{\dag}_{1}=\left(d^{\dag}_{1\uparrow},d_{1\downarrow}\right)$ and $\boldsymbol{\Psi}^{\dag}_{2}=\left(d^{\dag}_{2\uparrow},d_{2\downarrow}\right)$.

The lesser Green's function for both quantum dots can be calculated using

\begin{equation}
\label{G1(2)<}
\boldsymbol{G}^{<}_{2(1)}(\omega)=\boldsymbol{G}^{r}_{2(1)}\left( \omega \right)\boldsymbol{\Sigma}^{<}_{2(1)T}\left( \omega \right)\boldsymbol{G}^{a}_{2(1)}\left( \omega \right) \hspace{0.1cm},
\end{equation}

where the self-energy for QD2 can be written as

\begin{equation}
\label{self-e2}
\boldsymbol{\Sigma}^{<}_{2T}=\boldsymbol{\Sigma}^{<}_{L}+\boldsymbol{\Sigma}^{<}_{R}+\boldsymbol{t}^{\dag} \boldsymbol{G}^{r}_{1bare}\boldsymbol{\Sigma}^{<}_{S}\boldsymbol{G}^{a}_{1bare} \boldsymbol{t} \hspace{0.1cm},
\end{equation}

and that for QD1 as

\begin{equation}
\label{self-e1}
\boldsymbol{\Sigma}^{<}_{1T}=\boldsymbol{\Sigma}^{<}_{S}+\boldsymbol{t}^{\dag} \boldsymbol{G}^{r}_{2bare}\left(\boldsymbol{\Sigma}^{<}_{L}+\boldsymbol{\Sigma}^{<}_{R}\right)\boldsymbol{G}^{a}_{2bare} \boldsymbol{t} \hspace{0.1cm}.
\end{equation}

In eqs.~(\ref{self-e2}) and (\ref{self-e1}) $\boldsymbol{G}^{r (a)}_{1 bare}$ and $\boldsymbol{G}^{r (a)}_{2 bare}$ refers to the retarded (advanced) ``bare'' Green's function of the systems constituted by $SC-QD1$ and $L-QD2-R$ respectively. Besides, the self-energies $\boldsymbol{\Sigma}^{<}_{L(R)}$, $\boldsymbol{\Sigma}^{<}_{S}$ are obtained by using the fluctuation-dissipation theorem $\boldsymbol{\Sigma}^{<}_{i}=\boldsymbol{F}_{i}\left( \omega \right)\left[ \boldsymbol{\Sigma}^{a}_{i}-\boldsymbol{\Sigma}^{r}_{i}\right]$, where $i=L, R$ or $S$. With $\boldsymbol{F}_{i}$ the Fermi matrix given by \cite{Siqueira}

\begin{equation}
   \boldsymbol{F}_{i}\left(\omega\right) = \left(
      \begin{array}{cc}
        f_{i}  &  0    \\
            0   & \overline{f}_{i} \\
	\end{array}\right)  ,
\end{equation}

\noindent where the Fermi functions are defined as $f_{i}=f\left( \omega-\mu_{i} \right)$ and $\overline{f}_{i}=f\left( \omega+\mu_{i} \right)$ for $i=L, R$ and $f_{i}=f\left( \omega \right)$ for $i=S$.  The retarded (advanced) self-energies for the normal leads $L$ and $R$ can be written as follows

\begin{equation}
\label{GammaLR}
   \boldsymbol{\Sigma}^{r(a)}_{L(R)} = \mp\dot{\imath}\frac{\Gamma_{L(R)}}{2}\left(
      \begin{array}{cc}
        1  &  0    \\
        0   & 1 \\
	\end{array}\right) .
\end{equation}
For the SC lead, the corresponding retarded self-energy is
\begin{equation}
\label{GammaS}
   \boldsymbol{\Sigma}^{r}_{S} = -\dot{\imath}\frac{\Gamma_{S}}{2}\rho\left(\omega\right)\left(
      \begin{array}{cc}
        1  &  \frac{\Delta}{\omega}    \\
        \frac{\Delta}{\omega}   & 1 \\
	\end{array}\right)  ,
\end{equation}

\noindent where $\rho\left(\omega\right)$ is the modified BCS density of states, with the imaginary part accounting for the Andreev states within the gap. 
$\rho\left(\omega\right)=\left[-\dot{\imath}\hspace{0.05cm}\omega\hspace{0.05cm}\frac{\theta\left(\Delta-|\omega|\right)}{\sqrt{\Delta^{2}-\omega^{2}}}+|\omega|\hspace{0.05cm}\frac{\theta\left(|\omega|-\Delta\right)}{\sqrt{\omega^{2}-\Delta^{2}}} \right]$.

After a straightforward calculation, we obtain the normal current, $I_{LR}$, and the Andreev current, $I_{A}$ 

\begin{eqnarray}
I_{LR}&=&\frac{2e}{h}\int d\omega \hspace{0.1cm} T_{LR}\left(\omega\right)\left( f_{L}-f_{R}\right) \hspace{0.1cm} ,\nonumber \\
I_{A}&=&\frac{2e}{h}\int d\omega \hspace{0.1cm} T_{A}\left(\omega\right)\left( f_{L}-\overline{f}_{L}\right) \hspace{0.1cm} ,
\end{eqnarray}

\noindent the Fermi distributions $f_{L}=f_{L}\left( \omega-\mu_{L} \right)$ and $f_{R}=f_{R}\left( \omega-\mu_{R} \right)$ are the corresponding distributions for electrons in the leads L and R and $\overline{f}_{L}=f_{L}\left( \omega+\mu_{L} \right)$ is the Fermi distribution for holes. The chemical potentials $\mu_{L}$ and $\mu_{R}$ are fixed by the applied bias $V_{L}=+\frac{V}{2}$ and $V_{R}=-\frac{V}{2}$, while the superconductor chemical potential $\mu_{S}$ is set to zero \cite{note1}. $T_{LR}(\omega)$ is the transmission between normal leads ($L$ and $R$), given by $T_{LR}(\omega)=\widetilde{\Gamma}_{L}\widetilde{\Gamma}_{R}|G_{2,11}\left(\omega\right)|^{2}$. The Andreev transmission is given by \cite{Zhu} $T_{A}(\omega)=\widetilde{\Gamma}^{2}_{L}|G_{2,12}\left(\omega\right)|^{2}$. The index $j$ in the Green's function $G_{j,\alpha\beta}$ denotes the QD site while the labels $\alpha$, $\beta$ denote the elements of the matrix in the Nambu spinor space.
Notice that $T_{A}\left(\omega\right)$ is an even function of $\omega$ because the Andreev scattering involves both the particle and hole degrees of freedom.
The retarded Green's functions for QD2 and QD1 were presented in a previous work by some of the authors. \cite{AMCalle}

\section{Results}

In this section we discuss the transport properties at zero temperature ($T = 0$). In what follows we will consider  $\Gamma_{L}$ as the energy unit and $E_{F} = 0$. We have taken $D = 60\Gamma_{L}$ and the energy level of the embedded QD is fixed at $\epsilon_{2}=-3.5\Gamma_{L}$. We also consider the coupling between dots $t=0.02\Gamma_{L}$ and $\Gamma_{R}=\Gamma_{L}$. At the equilibrium and without interdot coupling, the Kondo temperature $T^{0}_{K}$ of QD2 is approximately $10^{-3}\Gamma_{L}$.

\begin{figure}[ht]
\centerline{\includegraphics[width=88mm,height=73mm,clip]{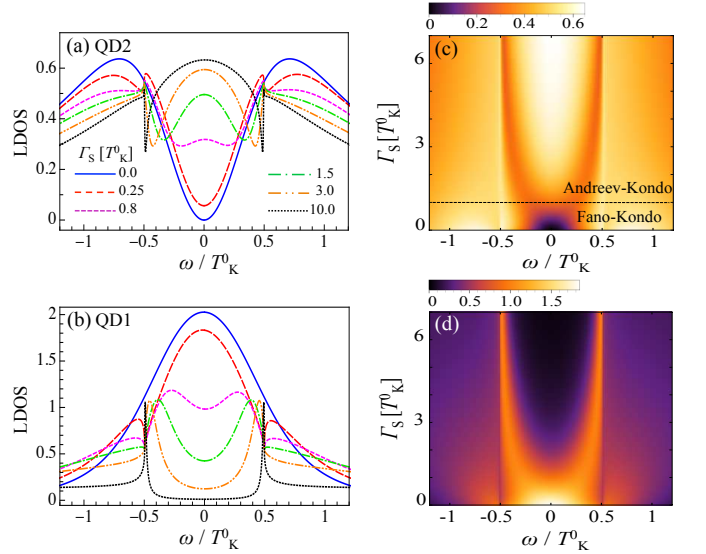}}
\caption{(Color Online) LDOS as a function of energy ($\omega$) for $\epsilon_{1}=E_{F}=0$, $\epsilon_{2}=-3.5\Gamma_{L}$, $\Delta=0.5 T^{0}_{K} $, $t=0.02\Gamma_{L}$ and $\Gamma_{R}=\Gamma_{L}$. a) and b) shows the LDOS for QD2 and QD1 respectively for several values of $\Gamma_{S}$: $\Gamma_{S}=0$ [solid (blue) line], $\Gamma_{S}=0.25 T^{0}_{K}$ [long dashed (red) line], $\Gamma_{S}=0.8T^{0}_{K}$  [dashed (magenta) line],  $\Gamma_{S}=1.5T^{0}_{K}$ [dotted dashed (green) line], $\Gamma_{S}=3T^{0}_{K}$  [double dotted dashed (orange) line]  and  $\Gamma_{S}=10T^{0}_{K}$ [dotted (black) line]. Panels c) and d) show the density plot of the LDOS for QD2 and QD1 respectively.}
\label{DOS}
\end{figure}

Fig.~\ref{DOS} shows the local density of states (LDOS) for QD2 [panel (a)] and QD1 [panel (b)] for several values of the coupling $\Gamma_{S}$ between QD1 and the SC lead ($\Delta=0.5\times T^{0}_{K}$ for all curves).
It is well-known that when $\Gamma_{S}=0$ and $\epsilon_{1}=0$, the LDOS of the embedded quantum dot (QD2) takes the form of a symmetric antiresonance [solid (blue) curve in panel (a)]. This antiresonance, which may be considered as a Fano-like structure (a suppression of the LDOS at and around $E_{F}$ in QD2), is caused by destructive interference between two conduction channels connecting leads $L$ and $R$: one
that passes straight through QD2, and another that `visits' QD1. In other words, the behavior of the LDOS of QD2 is a consequence of the Kondo effect modified by interference effects, which is clearly seen in the LDOS shown in Fig.~\ref{DOS}(a).
When the coupling between the SC lead and QD2 is turned on [long dashed (red) line],  two small kinks, located at the edges of the superconductor gap ($\omega=\pm\Delta$) develop, in addition to a transfer of spectral weight from outside the superconducting 
gap ($|\omega|/T_K^0 > \Delta$) to inside of it ($|\omega|/T_K^0 < \Delta$). 
As  $\Gamma_{S}$ further increases, and spectral weight keeps accumulating at and around $\omega=E_F$, the original broad Fano antiresonance splits in two narrower ones, which tend to localize at $\pm\Delta$ when $\Gamma_{S}>T^{0}_{K}$ and  the original Fano antiresonance centered around the Fermi energy becomes a resonance. 
Fig.~\ref{DOS}(b) displays the local density of states for the lateral quantum dot QD1. It can be clearly seen for the smallest finite values of $\Gamma_{S}$ ($\Gamma_{S}=0.25 T^{0}_K$, [long dashed (red) line]) that the LDOS shows a Lorentzian line-shape, centered around the Fermi energy, with two dips located at the edge of the superconducting gap $\omega=\pm\Delta$, which are due to  the presence of the SC lead. 
As $\Gamma_{S}$ increases, the original peak starts to split into two broad peaks [$\Gamma_{S}=0.8 T_K^0$, 
dashed (magenta) line], while for $\Gamma_{S}\geqslant T^{0}_{K}$ a pseudo gap is formed around the Fermi energy with a double-peak structure inside the gap originating from the superconducting proximity effect. This double peak structure becomes sharp for $\Gamma_{S}\gg T^{0}_{K}$ indicating the formation of the Andreev bound states. 
The change in the LDOS of QD1 from a single resonance to two Andreev resonances [Fig.~\ref{DOS}(b)] clearly characterizes the crossover between the dominant couplings at the QD, which occurs around $\Gamma_{S}  \approx T^{0}_{K}$. 
Both QD's LDOS, as a function of the energy and the coupling with the SC lead, are shown in Fig.~\ref{DOS}(c) (QD2) and (d) (QD1). We can clearly observe the formation and evolution of the Andreev bound states in QD1 [panel (d)] as $\Gamma_{S}$ increases. 
Figure \ref{DOS}(c) clearly shows the crossover region between the two regimes defined by the prevailing 
coupling in QD2: $\Gamma_{S} \approx T_{K}^0$ separates the regions where the Fano-Kondo 
($\Gamma_{S} < T_{K}^0$) and the Andreev-Kondo ($\Gamma_{S}>T_{K}^0$) effects are dominant.
The last assertion is general for $\Delta<<T^{0}_{K}$ and $\Delta>>T^{0}_{K}$. Initially, the continuum density of states in the $QD1$ is due to the coupling with $QD2$ where there is a  Kondo state, as the coupling between the $QD1$ and $S$-lead is turned on, Andreev bound states begin to form in $QD1$.  As $\Gamma_S$ increases the Andreev bound state are split in two peak centered around $\pm \Gamma_S/2$, however if $\Gamma_S > \Delta$ the Andreev bound states tend to localize at the edge of the superconducting gap $\pm \Delta$.

\begin{figure}[ht]
\centerline{\includegraphics[width=88mm,height=73mm,clip]{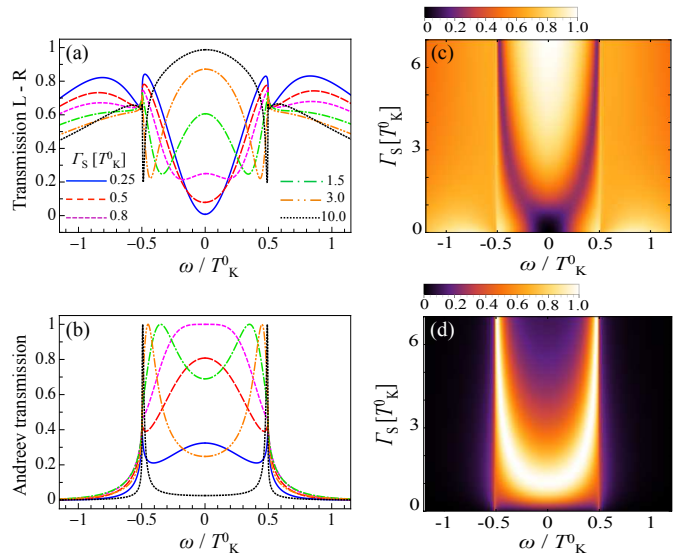}}
\caption{ (Color Online) Transmission from $L$ to $R$-lead and Andreev transmission for $\epsilon_{1}=E_{F}=0$, $\epsilon_{2}=-3.5\Gamma_{L}$, $\Delta=0.5 T^{0}_{K} $, $t=0.02\Gamma_{L}$ and $\Gamma_{R}=\Gamma_{L}$.  a) Transmission from $L$ to $R$-lead and b) Andreev transmission for several values of $\Gamma_{S}$: Blue solid line corresponds to $\Gamma_{S}=0.25T^{0}_{K}$, $\Gamma_{S}=0.5 T^{0}_{K}$ [long dashed (red) line], $\Gamma_{S}=0.8T^{0}_{K}$  [dashed (magenta) line],  $\Gamma_{S}=1.5T^{0}_{K}$ [dotted dashed (green) line], $\Gamma_{S}=3T^{0}_{K}$ [double dotted dashed (orange) line]  and  $\Gamma_{S}=10T^{0}_{K}$ [dotted (black) line]. Panels c) and d) shows the density plot for the normal transmission and Andreev transmission, respectively.}
\label{Transmission}
\end{figure}

\begin{figure}[ht]
\centerline{\includegraphics[width=54mm,height=71mm,clip]{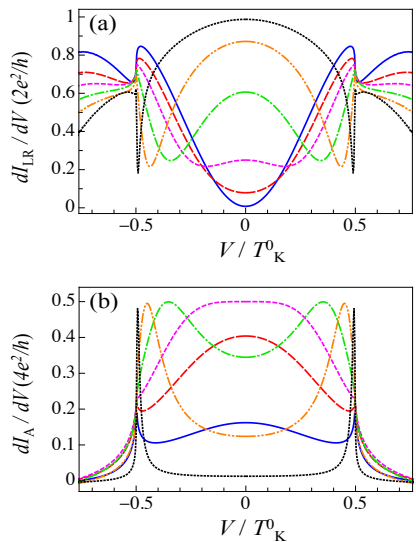}}
\caption{ (Color Online) a) Normal differential conductance and b) Andreev differential conductance for several values of $\Gamma_{S}$:  $\Gamma_{S}=0.25$ ([solid (blue) line], $\Gamma_{S}=0.5 T^{0}_{K}$ [long dashed (red) line], $\Gamma_{S}=0.8T^{0}_{K}$ [dashed (magenta) line],  $\Gamma_{S}=1.5T^{0}_{K}$ [dotted dashed (green) line], $\Gamma_{S}=3T^{0}_{K}$ [double dotted dashed (orange) line], $\Gamma_{S}=10T^{0}_{K}$ [dotted (black) line].}
\label{Diffcond}
\end{figure}

Figs.~\ref{Transmission}(a) and (b) display the transmission from normal left-right  leads and the  Andreev transmission, respectively. The transmission from $L$ to $R$ [Fig.~\ref{DOS}(a)] shows a very similar behavior to the QD2 LDOS. Like in that figure, the initial Fano antiresonance ($\Gamma_{S}=0$) becomes a resonance centered around Fermi energy as $\Gamma_{S}$ increases. This resonance is accompanied by two Fano antiresonances that tend to be positioned at $\omega=\pm\Delta$ when $\Gamma_{S}>T^{0}_{K}$. These resonances have identical shapes but an opposite sign of the imaginary asymmetry Fano parameter $q$.  Fig.~\ref{Transmission}(b) displays the Andreev transmission as a function of the energy for several values of the coupling $\Gamma_{S}$.  When $\Gamma_{S}$ is small ($\sim0.1T^{0}_{K}$) the hybridization between the system and the superconductor is negligible and the Andreev transmission presents two small peaks originating from the Andreev bound states of QD1.  When $\Gamma_{S}$ is gradually increased but still smaller than $T^{0}_{K}$, we can see a peak centered around zero energy that emerges for $\Gamma_{S}=0.25T^{0}_{K}$ and reach its largest value around $\Gamma_{S}\approx T^{0}_{K}$.  When $\Gamma_{S}>T^{0}_{K}$, the broad peak in the Fermi energy  splits and a double-peak structure appears. These two broad peaks evolve into two sharp peaks  as $\Gamma_{S}$ become larger than $T^{0}_{K}$. These peaks are associated  to the Andreev bound states. 
Figures \ref{Transmission}(c) and (d) show the density plots for normal and Andreev transmission, respectively. It is important emphasize the fact that in the crossover regions ($\Gamma_{S}\approx T^{0}_{K}$) the Andreev transmission reaches its maximum value as can be seen in Fig.~\ref{Transmission}(d).  Outside this region the formation of Andreev bound states take place.
As we pointed out in a previous work by some of the authors \cite{AMCalle}, it is important to highlight that the resonances exhibited in the Andreev transmission for $\Gamma_{S}>T^{0}_{K}$ [Fig.~\ref{Transmission}(b)] are centered in the same position as the Fano antiresonances in the transmission from $L$ to $R$ [Fig.~\ref{Transmission}(a)]. Therefore, this suggests, that there is a correlation between a Fano antiresonance in the normal transmission and the Andreev bound states. This correlation was called `Fano-Andreev effect' \cite{AMCalle} and as our current result show, it is robust against the introduction of correlation between electrons, like the Kondo effect.

Fig.~\ref{Diffcond} shows results for the normal differential conductance [Fig.~\ref{Diffcond}(a)] and the Andreev differential conductance [Fig.~\ref{Diffcond}(b)] as a function of the corresponding applied bias voltage. As expected, the normal differential conductance in Fig.~\ref{Diffcond}(a) resembles the LDOS of QD2 [Fig.~\ref{DOS}(a)]. For small values of $\Gamma_{S}$, the differential conductance shows two small kinks at $V_{LR}\approx T^{0}_{K}/2$ as a consequence of the hybridization with the S-lead. With further increasing of $\Gamma_{S}$, the initial Fano antiresonance develops into a peak at the Fermi energy and, as a consequence, a double Fano antiresonance structure emerges, which can be associated to the Andreev bound states. 
Results for the Andreev differential conductance can be seen in Fig.~\ref{Diffcond}(b). As $\Gamma_{S}$ increases, the Kondo resonant peak at zero voltage clearly appears and the amplitude of the conductance increases.  The zero-bias Andreev differential conductance starts to decrease after the crossover region ($\Gamma_{S}>T^{0}_{K}$). With further increase of $\Gamma_{S}$, the zero-bias Andreev differential conductance is strongly suppressed and the distance between the Andreev bound states peaks is augmented.

The results presented up to now were obtained for a small value of the superconducting gap $\Delta=T^{0}_{K}/2$. As Kondo temperatures for quantum dots are reasonably small energy scales (less than $1K$), in comparison with superconducting gaps of BCS superconductors (which can be as high as $10K$), it is of interest to analyze the large gap limit $\Delta \rightarrow \infty$, in which only the off-diagonal terms of the superconductor self-energy (\ref{GammaS})  are preserved tending to the static value  \cite{D} $-\Gamma_S/2$.  In this approximation the normal transmission, $T_{LR}$, and Andreev transmission, $T_{A}$, can be written as

\begin{widetext}
\begin{eqnarray*}
T_{LR}(\omega) &=&\widetilde{\Gamma}_L  \widetilde{\Gamma}_R \frac{  (\omega^2 (\omega^2 - \frac{\Gamma^{2}_S}{4} - \widetilde{t}^2)^2 + \frac{\widetilde{\Gamma}^{2}}
    {4} (\omega^2 - \frac{\Gamma^{2}_S}
      {4})^{2}}{((\omega^2 - \frac{\widetilde{\Gamma}^{2}}
         {4}) (\omega^2 - \frac{\Gamma^{2}_S}{4}) -  2 \hspace{0.05cm} \widetilde{t}^2 \omega^2 + \widetilde{t}^4)^2 + \widetilde{\Gamma}^2 \omega^2 (\omega^2 - \frac{\Gamma^{2}_S}{4} - \widetilde{t}^2)^2}  \hspace{0.15cm},      \\
 T_{A} (\omega) &=&\frac{\widetilde{\Gamma}^{2}_{L}\left(\frac{\Gamma_{S}}{2} \hspace{0.05cm} \widetilde{t}^{2}\right)^{2}}{\left( \left(\omega^{2}-\frac{\widetilde{\Gamma}^{2}}{4}\right)\left(\omega^{2}-\frac{\Gamma_{S}^{2}}{4}\right)-2 \hspace{0.05cm} \widetilde{t}^{2}\omega^{2}+\widetilde{t}^{4}\right)^{2}+\widetilde{\Gamma}^{2}\omega^{2}\left(\omega^{2}-\frac{\Gamma_{S}^{2}}{4} -\widetilde{t}^{2}\right)^{2}} \hspace{0.15cm},        
\end{eqnarray*}
\end{widetext}

\begin{figure}[ht]
\centerline{\includegraphics[width=57mm,height=75mm,clip]{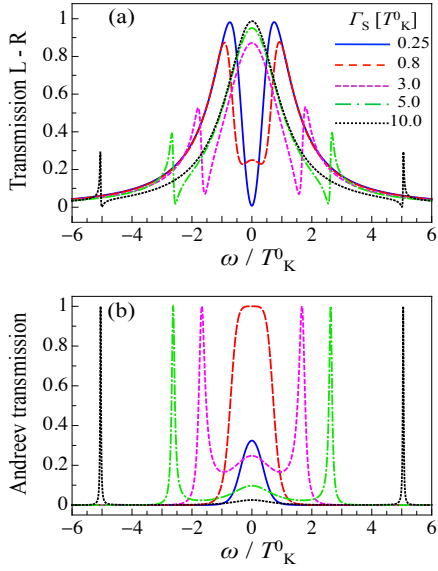}}
\caption{ (Color Online) a) Transmission from L to R-lead and b) Andreev transmission in the limit of large gap ($\Delta \rightarrow \infty$) for $\epsilon_{1}=0$, $\widetilde{\epsilon}_{2}\approx0$, $t=0.02\Gamma_{L}$ and $\Gamma_{R}=\Gamma_{L}$, for different  values of $\Gamma_S$:  $\Gamma_{S}=0.25$ [solid (blue) line] , $\Gamma_{S}=0.8 T^{0}_{K}$ [long dashed (red) line] , $ \Gamma_{S}=3T^{0}_{K}$ [dashed (magenta) line], $\Gamma_{S}=5T^{0}_{K}$ [dotted dashed (green) line)  and $\Gamma_{S}=10T^{0}_{K}$ [dotted (black) line]}
\label{farsubgap}
\end{figure}

\noindent and the results obtained with these expressions are displayed in Fig.~\ref{farsubgap}, where it is observed a similar behavior to the finite $\Delta$ case. We can clearly see that the original Fano antiresonance in the normal transmission [panel (a)] splits into two Fano line-shapes centered around $\pm \Gamma_S/2$. The results for the Andreev transmission [panel (b)] show that the small peak around the Fermi energy [$\Gamma_{S}=0.25T^{0}_{K}$, solid (blue) curve] increases up to unity for $\Gamma_{S}=0.8T^{0}_{K}$ [long dashed (red) curve], and then it splits into two peaks corresponding to the Andreev bound states.
As the coupling between QD1 and the SC lead is turned on, two Andreev bound states are induced in the quantum dot QD1, however if the separation between them is less than the Kondo temperature only one Andreev state is visible in the Andreev transmission. As the separation between the Andreev bound states is greater than the Kondo temperature, it is possible to observe both states in the Andreev transmission, after a crossover between both regimes. 

In order to get a better insight about the different transmission mechanisms occurring in this system, we take  the limit of small coupling between dots $\widetilde{t} \ll \widetilde{\Gamma}_{L} $ and $\Gamma_{S}\approx\widetilde{\Gamma}_{L}$. Then, the normal and  the Andreev transmission can be written approximately as  

\begin{eqnarray*}
T_{LR}(\omega)&\approx&\frac{\vert\varepsilon+q \vert ^2}{\varepsilon^2+1}\frac{1}{\xi^2+1} \hspace{0.1cm}, \\
T_{A}(\omega) &\approx& \frac{1}{\varepsilon^2+1} \hspace{0.1cm},
\end{eqnarray*}
with $\varepsilon=(\vert \omega \vert-\Gamma_S/2)/\eta$, $\eta=\tilde{t}^2/(\tilde{\Gamma}/2)$, $q=(1+2i)/4$ and $\xi=\omega/(\tilde{\Gamma}/2)$. From the above equations  we can see, on one hand, that the normal transmission is a  convolution of a Breit-Wigner line-shape (centered at the Fermi energy, and with width $T^{0}_{K}$) and Fano line shape (with a complex $q$-parameter, centered in $\pm \Gamma_S/2$ and with a width $\eta$). On the other hand, the Andreev transmission is a superposition of two Breit-Wigner lines shapes, centered at $\pm \Gamma_S/2$, and with a width $\eta$. It is worth noting that the Andreev bound states in the side attached quantum dot (QD1) get a width through the continuum of the embedded quantum dot (QD2), which is in the Kondo regimen.

\section*{Summary}
In this work, we have investigated the electronic transport properties through a T-shape double quantum dot coupled to a hybrid charge reservoir comprised of two normal and one SC lead (see Fig.~\ref{system}). The interplay between Kondo and proximity effect was studied considering a strong intradot Coulomb interaction $U$ in the embedded quantum dot. 
We found that the Kondo resonance is modified by the Andreev bound states which manifest themselves through Fano antiresonances in the local density of states and normal transmission. Hence, there is a correlation between Andreev bound states and Fano resonances that remains even in the presence of electron-electron interaction. 
We have also found that the dominant couplings at the quantum dots are characterized by a crossover region that defines the range where the Fano-Kondo and the Andreev-Kondo effect prevail in each QD. 
This conclusion is general for an arbitrary superconductor gap $\Delta$ ($\Delta<<T^{0}_{K}$ and $\Delta>>T^{0}_{K}$). Finally we found that the interplay between Kondo and Andreev bound states has a notable influence on the Andreev differential conductance.

\begin{acknowledgments}
We gratefully acknowledge the financial support from FONDECYT program Grants  No. 
3140053, 1140571 and 1151316,  CONICYT ACT 1204. G.B.M. and A.M.C. acknowledge financial support
from the NSF under Grants No.~DMR-1107994 and No.~MRI-0922811. We kindly thank E.~V. Anda for fruitful discussions.
We also want to thank J.~A. Ot\'alora from UTFSM for allowing us to use his computational facilities.
\end{acknowledgments}

\appendix
\section{Derivation of $T_{LR}$ and $T_{A}$ in the large gap limit ($\Delta\longrightarrow \infty$)}

As shown in our previous work \cite{AMCalle} the Green's functions for QD1 and QD2 can be determined from the following set of coupled equations
\begin{eqnarray}
\label{G1}
\mathbf{G}^{-1}_{1}\left( \omega \right) = \left[ \omega+\dot{\imath}\frac{\Gamma_{S}}{2}\rho\left(\omega\right) \right] \mathbf{\mathbb{I}} 
- &\epsilon_{1}&\mathbf{\sigma}_{z} 
- \dot{\imath}\frac{\Gamma_{S}}{2}\rho\left(\omega\right)\frac{\Delta}{\omega}\mathbf{\sigma}_{x} \nonumber\\
&-& \widetilde{t}^{2} \hspace{0.05cm} \mathbf{G}_{2 bare}\left(\omega\right),
\end{eqnarray}

\begin{equation}
\label{G2}
\mathbf{G}^{-1}_{2}\left( \omega \right) = \left[ \omega+\dot{\imath}\frac{\left(\widetilde{\Gamma}_{L}+\widetilde{\Gamma}_{R}\right)}{2}\right] \mathbf{\mathbb{I}} 
- \widetilde{\epsilon}_{2}\mathbf{\sigma_{z}} 
- \widetilde{t}^{2} \hspace{0.05cm} \mathbf{G}_{1 bare}\left(\omega\right),
\end{equation}
where $\mathbf{\mathbb{I}}$ is the identity matrix, $\mathbf{\sigma}_{x}$, $\mathbf{\sigma}_{z}$ denote the usual Pauli matrices, 
and $\rho\left(\omega\right)$ is the modified BCS density of states 
$\rho\left(\omega\right)=\left[-\dot{\imath}\hspace{0.05cm}\omega\hspace{0.05cm}\frac{\theta\left(\Delta-|\omega|\right)}{\sqrt{\Delta^{2}-\omega^{2}}}+|\omega|\hspace{0.05cm}\frac{\theta\left(|\omega|-\Delta\right)}{\sqrt{\omega^{2}-\Delta^{2}}} \right]$.
In eqs.~(\ref{G1}) and (\ref{G2}), $\mathbf{G}_{1 bare}$ and $\mathbf{G}_{2 bare}$ refer to the ``bare'' Green's function of the systems constituted by $SC-QD1$ and $L-QD2-R$, respectively. Using the Dyson equations 

\begin{equation}
\label{G1bare}
\left[\mathbf{G}^{r}_{1 bare}\right]^{-1}=\left[\mathbf{g}^{r}_{1}\left(\omega\right)\right]^{-1}-\mathbf{\Sigma}^{r}_{S},
\end{equation}

and
\begin{equation}
\label{G2bare}
\left[\mathbf{G}^{r}_{2 bare}\right]^{-1}=\left[\mathbf{g}^{r}_{2}\left(\omega\right)\right]^{-1}-\mathbf{\Sigma}^{r}_{L}-\mathbf{\Sigma}^{r}_{R},
\end{equation}
where the (unperturbed) Green's function for the side-attached quantum dot is given by 

\begin{equation}
\label{g1}
\left[\mathbf{g}^{r}_{1}\left(\omega\right)\right]^{-1}=
\left(
\begin{array}{cc}
  \omega-\epsilon_{1} & 0      \\
             0                       & \omega+\epsilon_{1}     
\end{array}
\right),
\end{equation}
and for the embedded quantum dot by 

\begin{equation}
\label{g2}
\left[\mathbf{g}^{r}_{2}\left(\omega\right)\right]^{-1}=
\left(
\begin{array}{cc}
  \omega-\widetilde{\epsilon}_{2} & 0      \\
             0                       & \omega+\widetilde{\epsilon}_{2}     
\end{array}
\right) .
\end{equation}

The retarded self-energies for the normal leads, $L$ and $R$, and for the SC lead, are given in equations (\ref{GammaLR}) and (\ref{GammaS}), in the main text.
In the large gap limit $\Delta \rightarrow \infty$, the modified BCS density of states $\rho\left(\omega\right)$ becomes in 
$\rho\left(\omega\right)=-\dot{\imath}\frac{\omega}{\Delta}$. As a consequence, in the large gap limit, $\Sigma_{S}$ is given by 

\begin{equation}
\label{GammaS_large}
   \mathbf{\Sigma}^{r}_{S} = -\frac{\Gamma_{S}}{2}\left(
      \begin{array}{cc}
        0  & 1    \\
        1  & 0 \\
	\end{array}\right) .
\end{equation}

Replacing equations (\ref{g1}), (\ref{g2}), (\ref{GammaLR}), and (\ref{GammaS_large}) in $\mathbf{G}^{-1}_{1}\left( \omega \right)$ and $\mathbf{G}^{-1}_{2}\left( \omega \right)$, we obtain the Green's functions for QD1 and QD2. It is 
straightforward to show that $G_{2,11}$ and $G_{2,12}$ are given by

\begin{equation}
\label{G211}
G_{2,11}\left(\omega\right)=\frac{1}{D}\left( \left(\omega+\widetilde{\epsilon}_{2}\right) + \dot{\imath}\frac{\widetilde{\Gamma}}{2}
- \frac{\widetilde{t}^{2}\left(\omega-\epsilon_{1}\right)}{\left(\omega^{2}-\epsilon^{2}_{1}\right)\-\left(\frac{\Gamma_{S}}{2}\right)^{2}} \right),
\end{equation}
and
\begin{equation}
\label{G212}
G_{2,12}\left(\omega\right)=-\dot{\imath} \hspace{0.05cm} \frac{1}{D} \left( \frac{\Gamma_{s}}{2}\frac{\widetilde{t}^{2}}{\left(\omega^{2}-\epsilon^{2}_{1}\right)-\left(\frac{\Gamma_{S}}{2}\right)^{2}}\right),
\end{equation}
where,

\begin{eqnarray}
\label{D}
D&=&\left(\omega^{2}-\widetilde{\epsilon}^{2}_{2}\right)-\left(\frac{\widetilde{\Gamma}}{2}\right)^2
-\frac{\widetilde{t}^{2}\left(\omega-\epsilon_{1}\right)\left(\omega-\widetilde{\epsilon}_{2}\right)}{\left(\omega^{2}-\epsilon^{2}_{1}\right)-\left(\frac{\Gamma_{S}}{2}\right)^2} \nonumber\\
&-&\frac{\widetilde{t}^{2}\left(\omega+\epsilon_{1}\right)\left(\omega+\widetilde{\epsilon}_{2}\right)}{\left(\omega^{2}-\epsilon^{2}_{1}\right)-\left(\frac{\Gamma_{S}}{2}\right)^2}
+\frac{\widetilde{t}^{4}}{\left(\omega^{2}-\epsilon^{2}_{1}\right)-\left(\frac{\Gamma_{S}}{2}\right)^2} \nonumber\\
&+&\dot{\imath}\hspace{0.05cm}\widetilde{\Gamma} \omega\left(1-\frac{\widetilde{t}^{2}}{\left(\omega^{2}-\epsilon^{2}_{1}\right)-\left(\frac{\Gamma_{S}}{2}\right)^2}  \right).
\end{eqnarray}

Finally, setting $\epsilon_{1}=\widetilde{\epsilon}_{2}=0$ in equations (\ref{G211}), (\ref{G212}), and (\ref{D}), we get the expressions for $T_{LR}$ and $T_{A}$ given in page 6: 

\begin{widetext}
\begin{eqnarray*}
T_{LR}(\omega) &=&\widetilde{\Gamma}_L  \widetilde{\Gamma}_R \frac{  (\omega^2 (\omega^2 - \frac{\Gamma^{2}_S}{4} - \widetilde{t}^2)^2 + \frac{\widetilde{\Gamma}^{2}}
    {4} (\omega^2 - \frac{\Gamma^{2}_S}
      {4})^{2}}{((\omega^2 - \frac{\widetilde{\Gamma}^{2}}
         {4}) (\omega^2 - \frac{\Gamma^{2}_S}{4}) -  2 \hspace{0.05cm} \widetilde{t}^2 \omega^2 + \widetilde{t}^4)^2 + \widetilde{\Gamma}^2 \omega^2 (\omega^2 - \frac{\Gamma^{2}_S}{4} - \widetilde{t}^2)^2}  \hspace{0.15cm},      \\
 T_{A} (\omega) &=&\frac{\widetilde{\Gamma}^{2}_{L}\left(\frac{\Gamma_{S}}{2} \hspace{0.05cm} \widetilde{t}^{2}\right)^{2}}{\left( \left(\omega^{2}-\frac{\widetilde{\Gamma}^{2}}{4}\right)\left(\omega^{2}-\frac{\Gamma_{S}^{2}}{4}\right)-2 \hspace{0.05cm} \widetilde{t}^{2}\omega^{2}+\widetilde{t}^{4}\right)^{2}+\widetilde{\Gamma}^{2}\omega^{2}\left(\omega^{2}-\frac{\Gamma_{S}^{2}}{4} -\widetilde{t}^{2}\right)^{2}} \hspace{0.15cm}.        
\end{eqnarray*}
\end{widetext}

\nocite{*}

\bibliography{apssamp}

\begin{references}
\bibitem{note1} It should be remarked that Crossed Andreev Reflections (CAR), whereby the 
reflected hole is scattered to the opposite lead, are also possibe. 
However, the current produced by CAR processes vanishes because of the particular symmetry of our system. 
Indeed, the CAR current generated by the L lead, given by 
$I_L^{CAR}= \frac{2e}{h}\int d\omega \Gamma_{L} \Gamma_{R} \hspace{0.05cm} |G^{r}_{2,12}|^{2} 
\left( f_{L}-\overline{f}_{R}\right)=0$, as $f_{L}=\overline{f}_{R}$. The same happens to the current 
generated by the R lead.
\end{references}

\end{document}